\documentclass[10pt]{iopart}
\usepackage{graphicx}
\usepackage{epsfig}
\usepackage{bm}
\usepackage{amssymb}
\usepackage{hyperref}

\begin{document}
	\title[Local magnetism, magnetic order and spin freezing in the `nonmetallic metal' FeCrAs]{Local magnetism, magnetic order and spin freezing in the `nonmetallic metal' FeCrAs}
	
	\author{B M Huddart$^1$, M T Birch$^{1,4}$, F L Pratt$^2$, 
		S J Blundell$^3$, D G Porter$^4$, S J Clark$^1$, W Wu$^5$, S R Julian$^{5,6}$, P D Hatton$^1$, T~Lancaster$^1$}
	
	\address{$^1$ Centre for Materials Physics, Durham University, Durham DH1 3LE, UK}
	\address{$^2$ ISIS Facility, Rutherford Appleton Laboratory, Chilton, Didcot 
		OX11 0QX, UK}
	\address{$^3$ Oxford University Department of Physics, Clarendon Laboratory, Parks
	Road, Oxford, OX1 3PU, UK}
	\address{$^4$ Diamond Light Source, Harwell Science and Innovation Campus, Didcot OX11 0DE, United Kingdom}
	\address{$^5$ Department of Physics, University of Toronto, Ontario M5S 1A7, Canada}
	\address{$^6$ CIFAR, Toronto, Ontario M5G 1Z8, Canada}
	\ead{benjamin.m.huddart@durham.ac.uk}
	
	\begin{abstract}
	We present the results of x-ray scattering and muon-spin relaxation ($\mu^{+}$SR) measurements on the iron-pnictide compound FeCrAs. 
	Polarized non-resonant magnetic x-ray scattering results reveal the 120$^\circ$ periodicity expected from the suggested three-fold symmetric, non-collinear antiferromagnetic structure.
	$\mu^+$SR measurements indicate a magnetically ordered phase throughout the bulk of the material below $T_\mathrm{N}$=105(5)~K.  There are signs of fluctuating magnetism in a narrow range of temperatures above $T_\mathrm{N}$ involving low-energy excitations, while at temperatures well below $T_\mathrm{N}$ behaviour characteristic of freezing of dynamics is observed, likely reflecting the effect of disorder in our polycrystalline sample. 
	Using density functional theory we propose a distinct muon stopping site in this compound and assess the degree of distortion induced by the implanted muon.
	\end{abstract}


		\ioptwocol
\section{Introduction}
Landau's Fermi liquid theory is perhaps the closest  condensed matter physics has  to a ``standard model''. 
However, metallic states not described by the Fermi liquid model are known to exist in many strongly correlated electron systems such as doped cuprates, quantum critical metals and disordered Kondo lattices \cite{RevModPhys.73.797}. Iron-pnictide superconductors also show
non-Fermi-liquid behaviour, most notably incoherent charge transport above a magnetic spin-density-wave transition, typically found at $T_{\mathrm{SDW}}~\approx~150$~K \cite{PhysRevLett.25.2215,PhysRevLett.101.107006,PhysRevLett.101.107007}.  Recent examples of materials which behave neither as a metal nor an insulator include high temperature superconducting cuprates \cite{Butch}, heavy fermion systems \cite{NatPhys5.465,PhysRevLett.104.186402}, iron-based chalcogenides \cite{PhysRevB.79.094521,PhysRevB.84.174506,PhysRevB.88.165110}, and oxyselenides \cite{PhysRevLett.104.216405,PhysRevB.89.100402,JPhysCM.28.453001}. Common to both cuprates \cite{Nature.518.179} and iron-based superconductors \cite{RevModPhys.87.855} is the emergence of high-temperature superconductivity from the suppression of the static antiferromagnetic order in their parent compounds.  The interplay between magnetism and unusual electronic transport is also evident in systems exhibiting a quantum critical point (QCP) arising from magnetic frustration \cite{Vojta} and in heavy-fermion superconductors, whose low temperature non-Fermi-liquid normal state most often derives from a nearby antiferromagnetic QCP \cite{Steglich}.

\begin{figure}[ht]
	\includegraphics[width=\columnwidth]{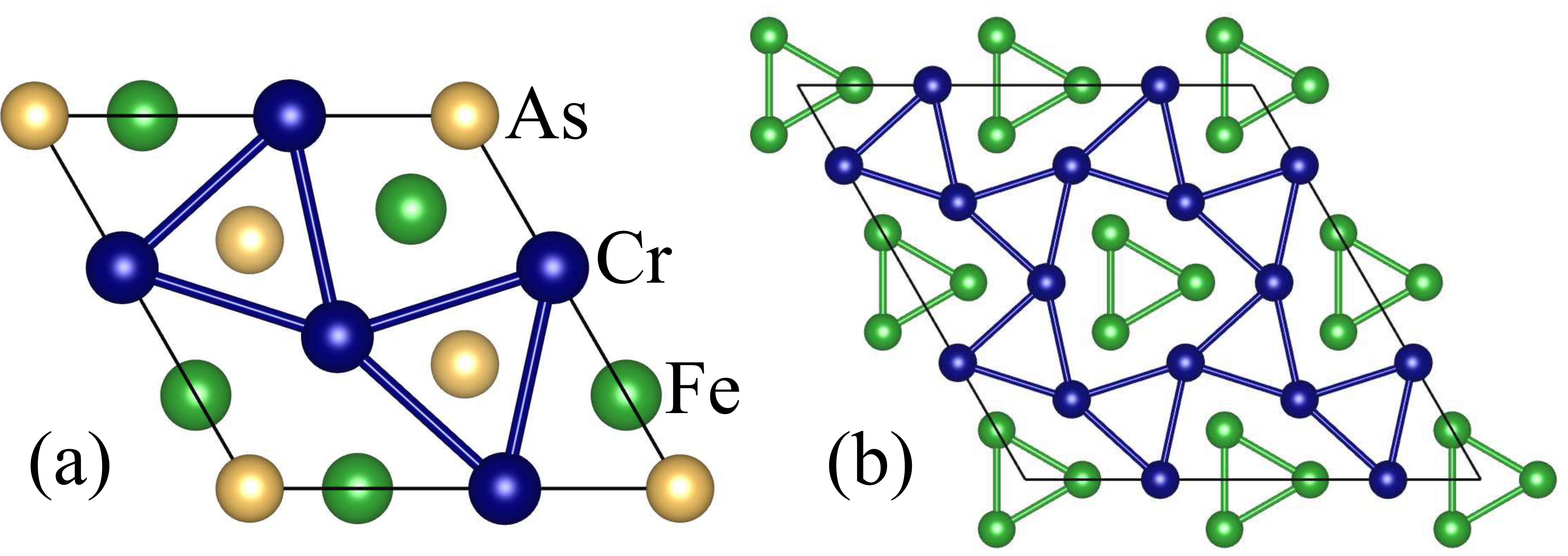}
	\caption{Structure of FeCrAs, showing (a) unit cell viewed along the $c$-axis and (b) Fe trimers and Cr kagome planes.}
	\label{figure1}
\end{figure}

The `nonmetallic metal' \cite{PhysRevB.89.125115} FeCrAs is a strongly correlated electron system, with local physics dominated by complex antiferromagnetism, magnetic frustration and charge fluctuations.  
Thermodynamic measurements show Fermi liquid-like specific heat; but resistivity has a non-metallic character \cite{0295-5075-85-1-17009}.  FeCrAs crystallizes in the hexagonal $P\bar{6}2m$ space group \cite{hollan,nylund,guerin} with Cr and Fe forming alternating two-dimensional lattices along the $c$-axis (see figure \ref{figure1}). 
Cr forms layers with the structure of a distorted kagome lattice where the  Cr--Cr distances are approximately constant. The Fe layers form a triangular lattice of three atom trimer units. The As is interspersed throughout both the Fe and Cr layers, as well as in between.  

Despite strong geometric frustration of the Cr sublattice, these moments are observed to order at $T_{\mathrm{N}}\approx 125$~K with an antiferromagnetic propagation vector of $\boldsymbol{k} =(1/3,1/3,0)$. Neutron diffraction \cite{doi:10.1139/P09-050} suggests that the ordered state has 
small moments between 0.6 and 2.2~$\mu_{\mathrm{B}}$ on the Cr atoms and, surprisingly, no detectable moment on the Fe sublattice. M\"{o}ssbauer spectra \cite{raincourt} show no magnetic splitting on the iron sites until far below $T_\mathrm{N}$, and even at the lowest temperatures a 
splitting of only (0.10~$\pm$~0.03)$\mu_B$ per Fe is observed. (Based on our findings, it seems likely that this tiny moment results 
from disorder in the polycrystalline samples used.)  Fe moments are also absent in x-ray emission spectroscopy, which found a suppressed Fe K $\beta^\prime$ line \cite{PhysRevB.84.100509}.  These studies show that, in contrast to iron-based pnictides, any static or dynamic Fe moment in FeCrAs is negligibly small at any temperature.

Theoretical studies have sought to explain the unusual metallic properties of FeCrAs. 
A model in which Heisenberg spins on the Cr kagome lattice couple either ferromagnetically or antiferromagnetically to Heisenberg spins at the average positions of a trimer results in a phase diagram consistent with the observed magnetic order \cite{arXiv:1105.3974}. A further theoretical study \cite{PhysRevB.84.104448} of the system addressed the lack of
static moments on the Fe atoms, by assuming that there are well-defined
local moments on the Fe sites and showed not only that the Fe sublattice would
play a role in stabilizing the magnetic order on the Cr sublattice, but that the electronic state on the Fe sublattice could be a $U(1)$ spin liquid phase that survives in the presence of the magnetic Cr sublattice. 
An alternative ``Hund's metal'' scenario \cite{PhysRevLett.103.147205,kotliar,stock} has been proposed involving localized moments coupling to itinerant electrons. 
Most recently \cite{stock} neutron scattering revealed that, despite the occurrence of magnetic ordering of the Cr sublattice with a mean field critical exponent at low temperatures, the dynamic response resembles that of an itinerant magnetic system at very high characteristic energies. The high energy scale of these excitations leaves open the possibility that the nonmetallic resistivity of FeCrAs at high temperature arises from scattering from spin and orbital fluctuations which are enhanced by magnetic frustration.  Even below $T_\mathrm{N}$, despite the probable absence of magnetic moments on the Fe sites, the fact that the ordered Cr moments are not fully polarized, but rather vary in size between 0.6 and 2.2 
$\mu_B$, allows for a scenario involving scattering of the conduction electrons from longitudinal fluctuations of the magnetic moment on the Cr sublattice, giving a possible mechanism for the continuation of the non-metallic resistivity to the lowest temperatures.

In this paper we investigate the unusual magnetic structure in FeCrAs by using polarized magnetic x-ray scattering to confirm the proposed magnetic state geometry. Implanted muons are sensitive to very small magnetic fields and so we have used these to probe the low-temperature order and dynamics in the magnetic state. We find that the sample is magnetically ordered throughout the bulk below $T_\mathrm{N}=105$~K, showing no evidence of phase separation.  However, we see evidence of low-energy spin excitations entering the time window of $\mu^+$SR below around 130~K, providing evidence of slow spin dynamics in FeCrAs that likely reflect fluctuations of the Cr moments. Below $T=20$~K a freezing of dynamics is observed, and this likely results from low-level disorder in the polycrystalline samples of this frustrated system.

\section{Experimental}

\begin{figure}[ht]
\begin{center}
	\includegraphics[width=7cm]{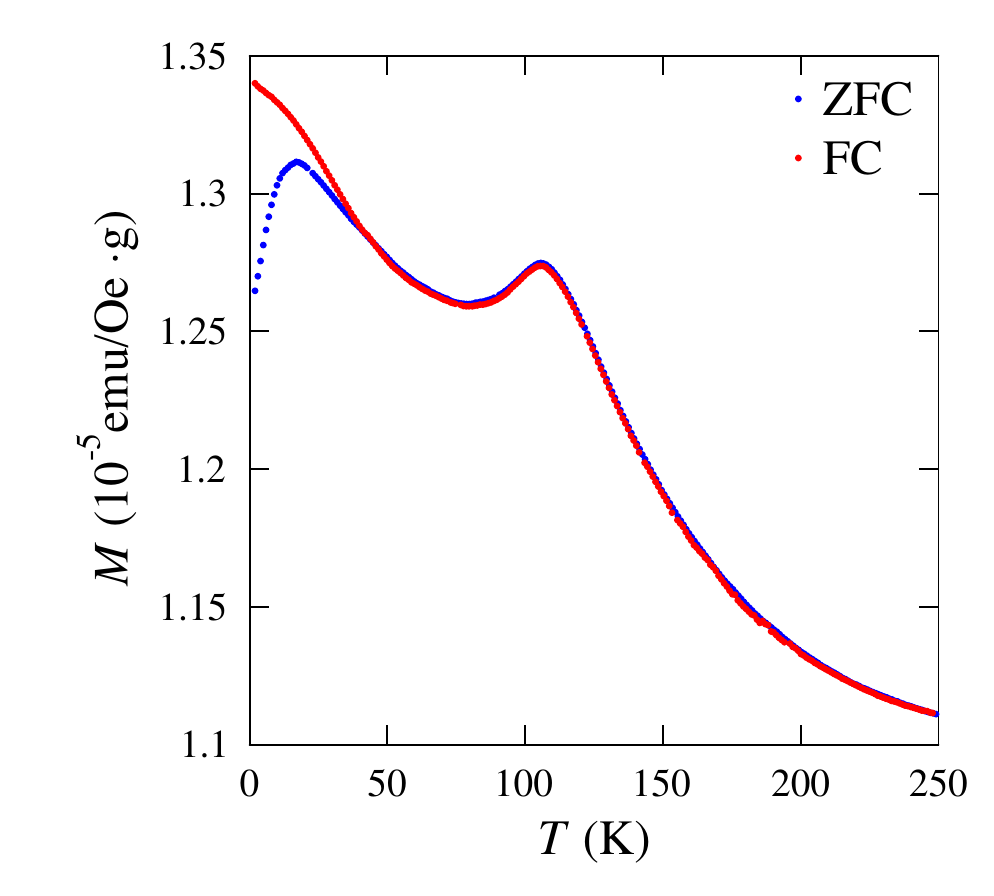}
	\caption{Magnetic susceptibility of a polycrystalline sample of FeCrAs measured in an applied field of 50~mT following zero-field cooling (ZFC) and field-cooling (FC) protocols.}
	\label{figure2}
\end{center}
\end{figure}

Polycrystalline and single crystal samples of FeCrAs were prepared as previously reported \cite{0295-5075-85-1-17009,wu_2011}.  Note that the polycrystalline sample was obtained by crushing high
quality single crystals, and then removing any ferromagnetic grains with a magnet. Such ferromagnetic grains are thought to contain iron inclusions. In subsequent magnetization measurements on these powder samples no 
magnetic hysteresis loops, that would be indicative of ferromagnetic domains, were resolved.
The magnetic properties of the polycrystalline sample were characterized using a Quantum Design MPMS system. The results of magnetic susceptibility are shown in figure~\ref{figure2}, where measurements following
zero-field cooling (ZFC) and field-cooling  (FC) protocols in an applied field of 50~mT are shown. Maxima in both ZFC and FC curves are seen at $T_{\mathrm{N}}=105$~K, corresponding to the transition to long range antiferromagnetic order. The ordering temperature has been observed to be quite sample dependent, with this value fairly typical of polycrystalline samples \cite{wu_2011}, (High-quality single crystals, such as the one measured in our x-ray study, typically show $T_{\mathrm{N}}\approx 125$~K.) The ZFC and FC curves are seen to diverge below 40~K, with the ZFC curve showing an additional maximum at $T=17$~K that is not seen in the FC curve. This divergence
is typical of a freezing of the magnetic moments at low temperature and reflects the fact that, after cooling in zero field
(ZF) to a region where the spins are statically frozen, we obtain a configuration that is less susceptible to magnetization in a small applied field compared to the more polarized state that occurs if the frozen state is formed in an applied field. The maximum in the ZFC curve has been attributed to a spin freezing transition \cite{wu_2011} occurring at a temperature that shows some variation between samples.

X-ray scattering measurements were made at the Diamond Light Source using beamlines I10 and I16 on the single crystal sample. Polarized x-ray scattering exploits the dependence of the scattering cross-section on the direction of the magnetic moment with respect to the incident and scattered polarization of the x-ray beam. Azimuthal measurements, where the sample is rotated through $\Psi$ around the scattering vector, change the magnetic moment direction with respect to the x-ray beam geometry, and consequently alter the scattering cross-section, allowing complex magnetic structures to be investigated.  Normally, measurements are taken exploiting the resonant enhancement of the magnetic scattering close to the atomic absorption edges.  This has the additional benefit of providing atom specificity to the observed magnetic signal.  Unfortunately we were not able to observe any magnetic resonances at either the Fe L or As L edges or the Cr K edges and hence we had to undertake non-resonant magnetic scattering.  This technique is not atom specific, but does have the advantage of being easier to interpret and model.

Muon-spin relaxation ($\mu^{+}$SR) measurements \cite{dalmas,steve_review}
were made on a polycrystalline sample  using the GPS instrument at the Swiss Muon Source (S$\mu$S), Paul Scherrer Institut, Villigen, Switzerland and at the ISIS facility, Rutherford Appleton Laboratory, UK, using the EMU instrument.
Measurements were made at S$\mu$S in a He cryostat where the sample was wrapped in Ag foil (thickness 25$~\mu$m) and mounted on a fork to minimize any background contribution. Measurements at ISIS were made using a closed-cycle refrigerator where the sample was mounted on a Ag plate to ensure good thermal contact.  

In a $\mu^{+}$SR experiment spin polarized muons are implanted into the sample. The quantity of interest
is the asymmetry $A(t)$, which is proportional to the spin polarization of the muon ensemble.  A quasistatic local magnetic field will result in coherent precession of the muon spin polarization.  The frequency of these oscillations is proportional to the local magnetic field $B$ at the muon site via $\nu=\gamma_{\mu} B / 2\pi$ where $\gamma_{\mu}$ is the muon gyromagnetic ratio ($=2\pi \times 135.5~\mathrm{MHz} ~\mathrm{ T}^{-1}$).  For a distribution of magnetic fields the muon spins will each precess at a different frequency, resulting in relaxation of the asymmetry.  When dynamics are present in the fast fluctuation limit \cite{dalmas, PhysRevB.20.850}, the relaxation rate is expected to vary as $\lambda \propto  {\Delta}^{2} \tau$, where $\Delta =\sqrt{{\gamma}_{\mu}^{2} \langle(B-\langle B \rangle)^{2} \rangle }$ is the second moment of the local magnetic field distribution at the muon site and $\tau$ is the correlation time.
At a continuous muon source (such as S$\mu$S) the 1 ns time resolution allows us to observe rapid relaxation and oscillation frequencies corresponding to large internal magnetic fields, but the beam-borne background makes the data measured at longer times uncertain.  Measurements made at a pulsed source (such as ISIS) are complementary because although the finite pulse width hinders our ability to resolve oscillations or relaxation rates $\gtrsim$ 10 MHz, the longer time window of 32 $\mu$s allows us to study slow spin dynamics.
 
\section{X-ray measurements}
Typically, magnetism in 3$d$ transition metals is examined with x-rays by tuning to the L$_3$ and L$_2$ edges of the magnetic atom, where the magnetic scattering is dramatically resonantly enhanced.  However, our initial measurements, made using the RASOR beamline at I10, found no magnetic satellites at $(1/3, 1/3, 0)$-type positions, which would be expected from the proposed magnetic structure around either the Fe L$_{2,3}$ or the As L$_{2,3}$ absorption edges.  

No wavevectors of the magnetic reflections of FeCrAs are accessible within the Ewald sphere at the softer Cr L$_{3}$ edge (574~eV). However, tuning to the K-edge offers an indirect method to observe magnetism via quadrupolar transitions in the pre-edge region, undergoing transitions into the magnetically active Cr-3$d$ orbital, providing an atomically selective probe of the magnetic moment residing on the Cr ions. A study of the azimuthal dependence of polarized magnetic x-ray scattering at the Cr K-edge (5.989~keV) was therefore undertaken using the I16 beamline.

\begin{figure*}[ht]
	\centering
	\includegraphics[width=0.9\textwidth]{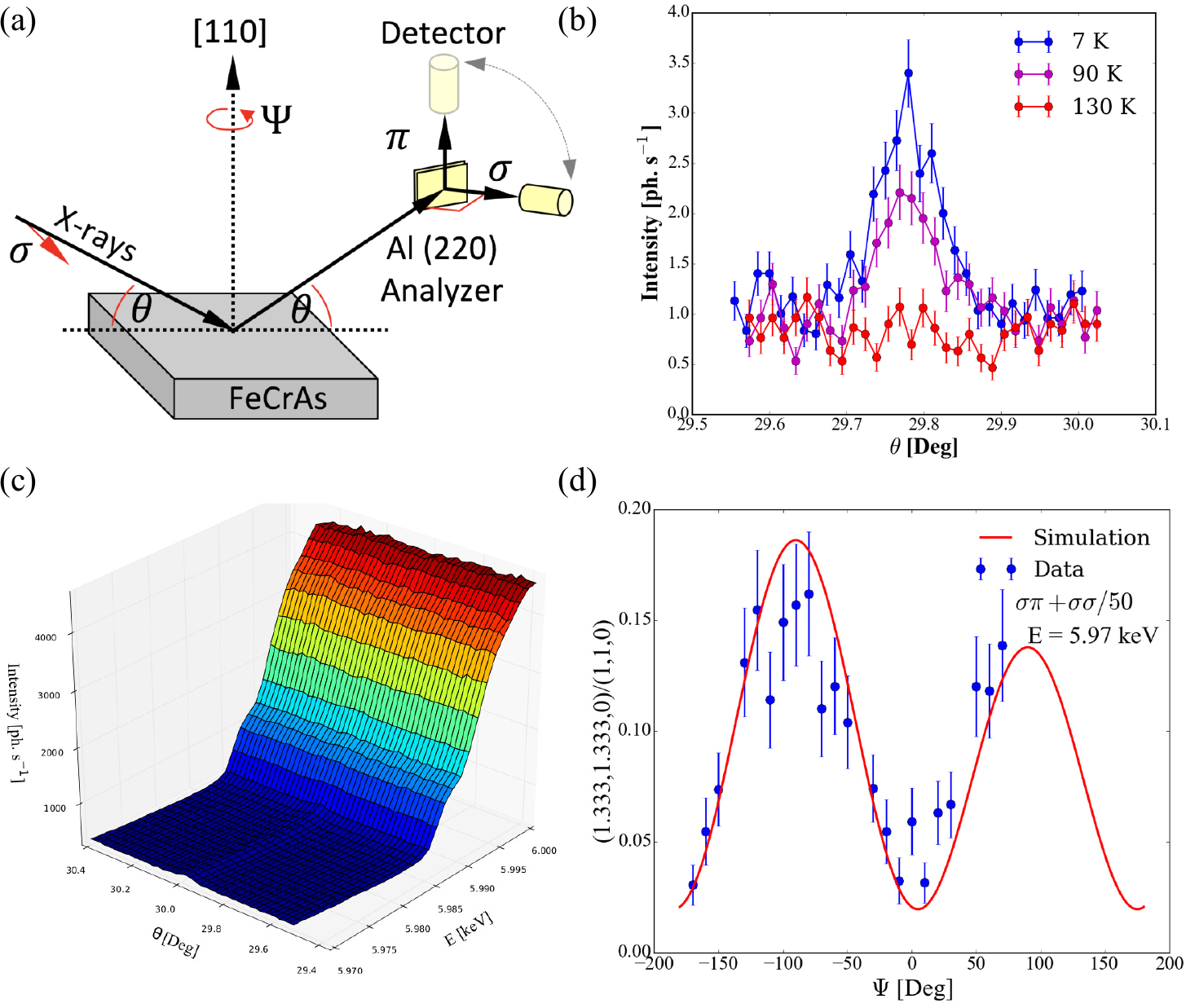}
	\caption{(a) Scattering geometry of the x-ray measurements. Incoming x-rays are linearly polarized in the $\sigma$ channel, scatter from the sample, and then pass through a $\sigma$ or $\pi$ analyzer before being detected. (b) Scattered intensity of the $(4/3,4/3,0)$ magnetic reflection as a function of $\theta$ for temperatures below and above $T_\mathrm{N}$, at $E$=5.97~keV.  The peak vanishes above $T_\mathrm{N}$, confirming its origin from the magnetic order. (c) Variation of the scattered intensity as a function of $\theta$ plotted at a variety of energies just below the Cr K absorption edge.  The edge is observed just below 6 keV together with a pre-edge feature at $\approx 5.99$~keV.  Magnetic scattering is observed at all energies in the range $5.97-5.99$~keV. (d) Normalized intensity of the $(4/3, 4/3, 0)$ magnetic reflection as a function of azimuthal angle, $\Psi$ [azimuthal reference (001)]. The expected non-resonant, spin-only, magnetic scattering intensity from the proposed non-collinear structure is shown alongside the data.}
	\label{figure3}
\end{figure*}

A single crystal sample ($\sim 2 \times 2 \times 0.5$ mm$^3$) with a polished $(110)$ surface was mounted in the beamline cryostat. The spectral $(220)$ Bragg peak was found which was both a singlet and very sharp (FWHM 0.06$^{\circ}$). Off-axis Bragg peaks such as the $(222)$, $(221)$ and $(4,-2, 0)$ were used to obtain an orientation matrix. X-rays incident on the sample are linearly polarized horizontally to the sample scattering plane and the polarization of the diffracted beam was linearly analyzed by rotating the scattering plane of an Al (220) single crystal as shown in figure \ref{figure3}(a). An Al $(220)$ single crystal was used as an analyzer as it has a suitable $d$-spacing to reflect x-rays at 90$^{\circ}$ (2$\theta$) at 6.1~keV, which is close to the Cr K-edge. The analyzer crystal was aligned on the sample $(110)$ reflection at 5.97~keV where a leakage signal of $\sigma\pi=\sigma\sigma/50$ was measured. Scans were made in the $\sigma$ - $\pi^{\prime}$ channel to maximize the magnetic signal and minimize charge scattering [see figure \ref{figure3}(a)]. 

Measurements were made after cooling to $T~=~7.5$~K, well below the N\'{e}el temperature. Magnetic reflections were found at 5.98 keV in the pre-edge region of the Cr K-edge with a propagation vector of $(1/3,1/3,0)$, 
together with much stronger magnetic peaks at $(2/3, 2/3, 0)$ and $(4/3, 4/3, 0)$. These peaks correspond to magnetic satellites with wavevectors of $(1/3, 1/3, 0)$ surrounding the $(110)$ Bragg reflection. The peaks were observed up to $T=110$~K, but were absent above $T_\mathrm{N}$ [see figure \ref{figure3}(b) for the analogous $(4/3, 4/3, 0)$ peak at $E=5.97$~keV], confirming their origin from the magnetic order. However, energy scans at constant wavevector showed no resonances close to the Cr K-edge. Indeed, the magnetic scattering signal was observed with a higher signal/background further away from the Cr K-edge at 5.97~keV.  This is shown in figure \ref{figure3}(c) where scans as a function of the scattering angle $\theta$ are displayed, showing the variation of the intensity as a function of the incident energy.  The small magnetic scattering peak, which is observed at all energies below the Cr absorption edge, can be seen at $\theta \approx 30^{\circ}$ [see also figure \ref{figure3}(b)].  The magnetic scattering displays little variation away from edge, being observable at all incident energies, with less fluorescence background and sample absorption below the Cr K-edge.  We therefore ascribe this scattering to non-resonant magnetic scattering.

The intensity of the $(4/3,4/3, 0)$ magnetic satellite was measured as a function of azimuthal angle by rotating the sample through $\Psi$ [with azimuthal reference (001)], as shown in figure~\ref{figure3}(d). The intensity was normalized with respect to the intensity of the $(110)$ Bragg peak recorded at each azimuth. The magnetic scattering signal varies as a function of azimuthal angle, displaying a maximum $\Psi=-100^{\circ}$ and a minimum around $\Psi=0^{\circ}$. The total amplitude for non-resonant magnetic scattering is given by \cite{Hill,Blume1985,BlumeGibbs}
\begin{equation}\label{equation1}
f^{(\mathrm{mag})}_\mathrm{non-res}=ir_0(\hbar \omega /mc^2)f_\mathrm{D} \left[\frac{1}{2} {\bf L} \left(\textbf{Q} \right) \cdot \textbf{A}+ {\bf S} \left(\textbf{Q} \right) \cdot \textbf{B} \right],
\end{equation}
where ${\bf L} \left(\textbf{Q} \right)$ and ${\bf S} \left(\textbf{Q} \right)$ are the Fourier transforms of the atomic orbital and spin magnetic densities, respectively.  In equation \ref{equation1}, $\textbf{A} =2(1 - {\bf \hat{k}} \cdot  {\bf \hat{k}'})({ \hat{\boldsymbol \epsilon}}' \times {\hat{\boldsymbol \epsilon}})-({ \hat{\bf k}} \times {\hat{\boldsymbol \epsilon}})({ \hat{\bf k}}\cdot {\hat{\boldsymbol \epsilon}'})+({ \hat{\bf k}}' \times {\hat{\boldsymbol \epsilon}'})({ \hat{\bf k}}' \cdot {\hat{\boldsymbol \epsilon}})$ and $\textbf{B} =({ \hat{\boldsymbol \epsilon}}' \times {\hat{\boldsymbol \epsilon}})+({ \hat{\bf k}}' \times {\hat{\boldsymbol \epsilon}'})({ \hat{\bf k}}' \cdot {\hat{\boldsymbol \epsilon}})-({ \hat{\bf k}} \times {\hat{\boldsymbol \epsilon}})({ \hat{\bf k}}\cdot {\hat{\boldsymbol \epsilon}'})-({ \hat{\bf k}'} \times {\hat{\boldsymbol \epsilon}'}) \times ({ \hat{\bf k}} \times {\hat{\boldsymbol \epsilon}})$, where $\boldsymbol \epsilon$ and ${\boldsymbol \epsilon}'$ are the incident and scattered polarization vectors respectively,
and \textbf{k} and $\textbf{k}'$ are the incident and scattered wave
vectors respectively. The expression also makes use of the wave-vector transfer $\textbf{Q} = \textbf{k}'-\textbf{k}$ 
and the Debye-Waller factor $f_\mathrm{D}$.
Considering spin-only scattering, we have used equation \ref{equation1} with $\textbf{L}(\textbf{Q})=\textbf{0}$ and simulate the expected scattering intensity from the proposed non-collinear antiferromagnetic magnetic structure \cite{doi:10.1139/P09-050} (shown in figure \ref{figurea2}, in the Appendix). Scaling was made by normalizing the sum of the squared intensity and the small $\sigma\sigma$ leakage component (=$\sigma\sigma/50$) seen in a comparable scan of the (110) charge reflection. The results of this model [shown in figure \ref{figure3}(d)] agree with the data within the experimental uncertainties.
We conclude that these results are consistent with the proposed three-fold symmetric, non-collinear antiferromagnetic structure.  As we have only carried out an azimuthal scan around a single direction, we cannot systematically rule out all other possible in-plane structures.  Despite this limitation, our data demonstrate that the magnetic structure suggested by neutron diffraction measurements \cite{doi:10.1139/P09-050} successfully describes the non-resonant magnetic scattering from FeCrAs for this experimental geometry. 

The magnetic structure suggested by neutron diffraction has ordered moments occurring only on the Cr atoms, with magnitudes between 0.6 and 2.2 $\mu_{\mathrm{B}}$ and no measurable moments on the Fe atoms \cite{doi:10.1139/P09-050}.  First principles electronic band structure calculations performed on paramagnetic FeCrAs show a significant density of states (DOS) close to the Fermi level originating from both Cr and Fe atoms \cite{PhysRevB.89.125115}.   However, the As atoms  contribute very little to the total DOS at $E_\mathrm{F}$, which probably explains the lack of magnetic scattering at the As L edges. The absence of a measurable signal at the Fe L edges suggests little hybridization between the Cr and Fe atoms. The lack of a resonant signal at the Cr K-edge is easier to explain. The K-edge corresponds to a 1s-4p transition and would require a quadrupolar transition into the partially occupied 3d shell. However, the resonant enhancement at the K-edge is much less than at the L edges, making any magnetic signal more difficult to observe. For there to be a resonant peak below the Cr K-edge would require a difference between the spin up and spin down channels to give an antiferromagnetic signal at $(1/3, 1/3, 0)-$type positions.  (To the best of our knowledge, spin-polarized electronic structure calculations of the expected DOS from the magnetic structure of FeCrAs have not yet been carried out.) There would also need to be a sharp peak in the DOS close to $E_\mathrm{F}$ from the Cr atoms to give a resonant enhancement.  
\section{Muon-spin spectroscopy}

\begin{figure}[h]
	\centering
	\includegraphics[width=\columnwidth]{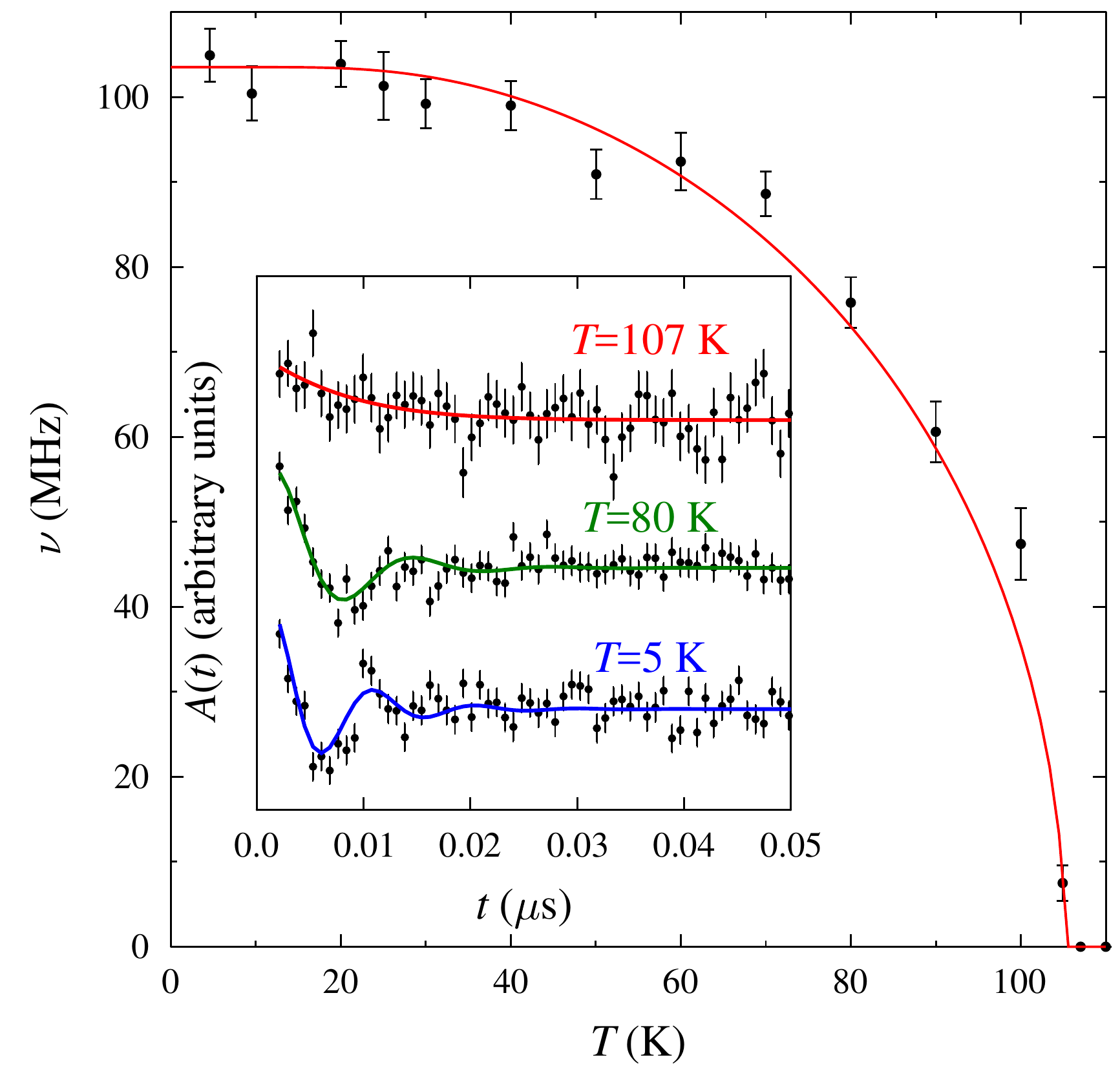}
	\caption{Temperature dependence of the precession frequency $\nu$ observed in zero-field $\mu^+$SR measurements made at S$\mu$S with an $S=3/2$ mean-field fit. Inset: Data and fits to the first 0.05 $\mu$s of the asymmetry spectra measured at several temperatures.}
	\label{figure4}
\end{figure}

Example zero-field (ZF) $\mu^+$SR measurements made at S$\mu$S on a polycrystalline sample of FeCrAs are shown in figure \ref{figure4}.  
Below $T=105$~K, we observe rapid oscillations in the early times of the asymmetry spectra, characteristic of quasistatic long range magnetic order (LRO) at the muon stopping site.  
We observe oscillations at a single frequency, which suggests a single magnetically distinct muon stopping site in the material or, more likely for this system, a single crystallographically distinct site experiencing a range of fields centred on the field corresponding to muon precession frequency.  The variance of local fields experienced at this site will then contribute to the observed rapid relaxation of the oscillations.
To parametrize the oscillations, we fit the first 0.05~$\mu$s of the spectra to a function
\begin{equation}
A(t)=A_{1} e^{-\lambda_\mathrm{f} t} \cos(2 \pi \nu t+\phi) + A_{2}, 
\label{equation2}
\end{equation}
where $A_1$ is the asymmetry due to muons precessing, whereas $A_2$ reflects muon-spin components aligned parallel to the internal magnetic field along with any implanting the sample holder.  
%(We note the absence of an additional purely relaxing component, which might be expected to arise from any fluctuating Fe spins.) 

The amplitudes $A_1$ and $A_2$ were allowed to vary, but we consistently found $A_1 \approx 2 A_2$ for all $T<105$~K. This is expected in a (quasistatic) magnetically ordered polycrystalline sample, where 2/3 of the initial muon polarization is expected to initially lie perpendicular to the local magnetic field and contribute to the precession signal, while 1/3 is expected to lie along the local field direction and contribute a constant. This suggests that the sample is magnetically ordered throughout its bulk with all muon sites subject to the same magnetic field distribution. Additionally in this measurement, where the sample is mounted on a fork, there is very little background contribution from the sample holder.  To ensure good fits we had to introduce a fixed phase offset of $\phi=-62^{\circ}$, determined from a low temperature fit. 
Non-zero phases, such as those observed here, have been observed in other $\mu^+$SR experiments \cite{Major1986,RevModPhys.69.1119} and often result from an asymmetric magnetic field distribution or, likely in this case, from difficulty in resolving features at early times in the spectra.  
The relaxation rate $\lambda_\mathrm{f}$ was found to be relatively large and approximately independent of temperature and so was fixed at its average value $\lambda_\mathrm{f}=173~\mu \mathrm{s}^{-1}$.  

In light of recent neutron diffraction measurements, where mean-field-like behaviour was reported \cite{stock}, the extracted muon precession frequency was fit according to mean-field theory.  An $S=3/2$ fit, appropriate for Cr$^{3+}$, yielded $T_{\mathrm{N}}=105(5)~\mathrm{K}$ and $\nu_\mathrm{sat}=104(1)~\mathrm{MHz}$ for the saturation magnetization.  This value of $T_{\mathrm{N}}$ is somewhat lower than the transition temperature of 125~K obtained from measurements of magnetic susceptibility and heat capacity on single crystals \cite{0295-5075-85-1-17009}, but is in good agreement with magnetic susceptibility measurements made on our polycrystalline sample.

Above $T_\mathrm{N}$ the spectra comprise a fast-relaxing component ($\lambda \approx 300$ MHz for $T=110$~K) and a component that relaxes very slowly (i.e.\ it is approximately constant on the time scale of the oscillations).  On cooling from $\approx 130$~K to $\approx 80$~K we observe a significant increase in the amplitude of the fast-relaxing component (accompanied by a drop of the other component).  (Below we attribute this behaviour to slow fluctuations entering the muon time window in this regime.)

\begin{figure*}[h]
	\centering
	\includegraphics[width=0.8\textwidth]{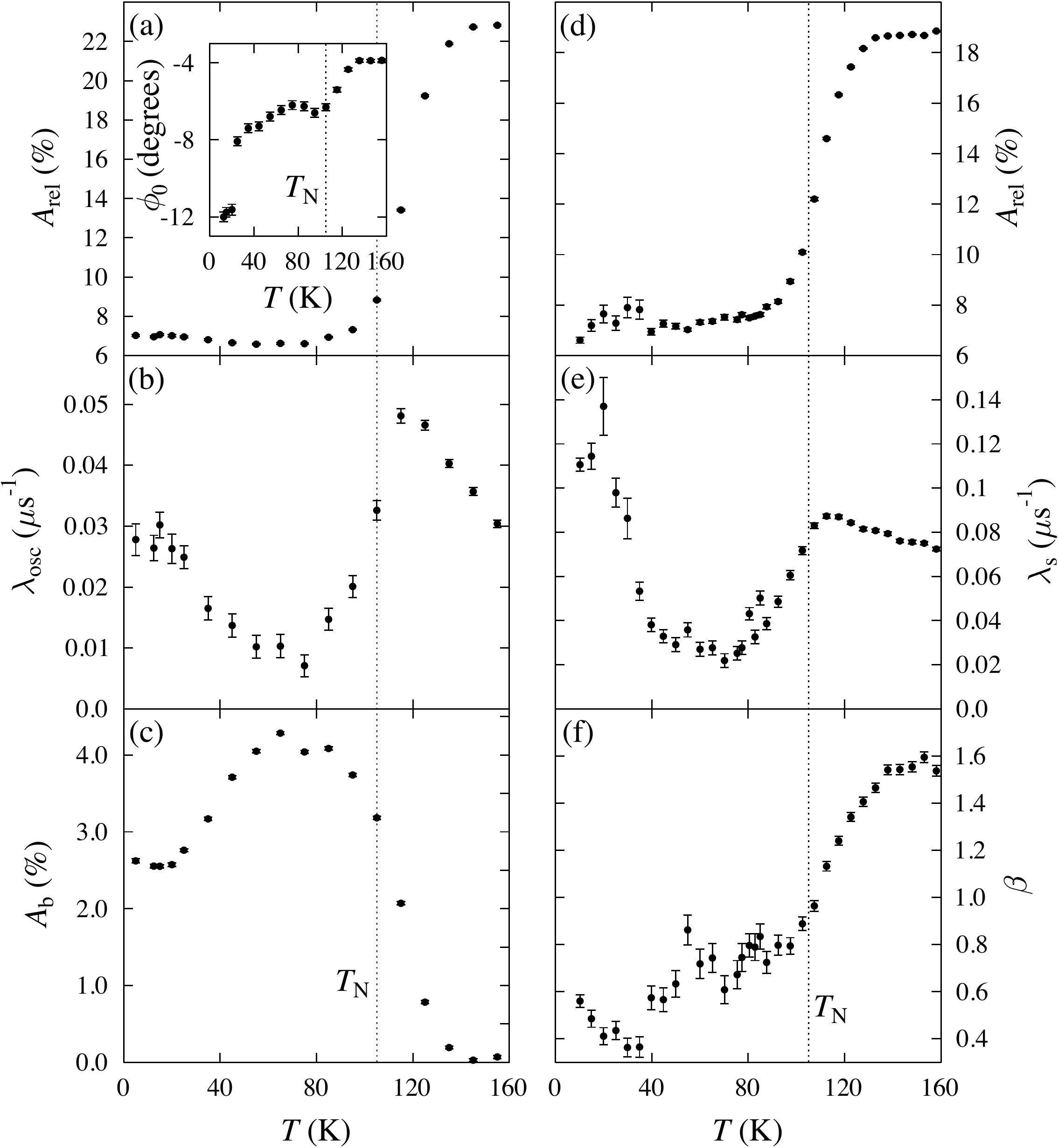}
	\caption{Results of fitting equation \ref{equation3}  to the wTF results, showing the temperature evolution of (a) relaxing asymmetry $A_{\mathrm{rel}}$, (b) relaxation parameter $\lambda_\mathbf{osc}$ and (c)  baseline asymmetry $A_{\mathrm{b}}$.  The temperature dependence of the phase $\phi_0$ is shown inset in (a). Temperature dependence from ZF measurements of (d) relaxing asymmetry $A_\mathrm{rel}$, (e) relaxation parameter  $\lambda_\mathrm{s}$ and (f) lineshape parameter $\beta$.}
	\label{figure5}
\end{figure*}

To investigate the possibility of multiple magnetic phases or other contributions to the muon depolarization, measurements were made at ISIS in an applied weak transverse field (wTF) of magnitude $B_{0}=2$~mT.  This field is small compared to the internal magnetic field below 100~K and so only those muons not in sites subject to the large internal field will be seen to precess at a frequency $\nu_{0} = \gamma_{\mu}B_{0}/2 \pi=0.28$~MHz. 
As seen above, the asymmetry from muons with spin components perpendicular to the internal magnetic field decays within 0.05~$\mu$s, so is not resolvable with the time resolution available at ISIS. 
In the absence of dynamic fluctuations we therefore expect the 2/3 of the muon spin components perpendicular to the internal fields to be quickly dephased from the spectrum, with the remaining 1/3 fixed along the large, static internal field direction and contributing a constant offset. This gives rise to the so-called 1/3-tail, which can only be relaxed by dynamics in the local magnetic field distribution. The spectra were therefore fitted to a function of the form
\begin{equation}
	A(t)=A_{\mathrm{rel}}e^{-\lambda_\mathrm{osc} t}\cos(2 \pi \nu_{0} t + \phi_0)+A_{\mathrm{b}},
	\label{equation3}
\end{equation}
where the phase $\phi_0$ was varied to ensure good fits.   
The resulting temperature dependence of the relaxing asymmetry $A_{\mathrm{rel}}$, the relaxation $\lambda_\mathrm{osc}$ and the baseline asymmetry $A_{\mathrm{b}}$ is shown in figure \ref{figure5}.

On crossing $T_{\mathrm{N}}$ from above, the oscillating component of the asymmetry $A_{\mathrm{rel}}$ [figure \ref{figure5}(a)] drops slowly by $\approx 15$\%, decreasing from from 22\% at 130~K to 7\% at 80~K, and does not vary significantly at lower temperatures.  (This behaviour of $A_{\mathrm{rel}}$ is consistent with the asymmetry measured at S$\mu$S in this region described above.)  The width in temperature of the change in $A_{\mathrm{rel}}$ is much broader than that typically observed in magnetic ordering transitions \cite{dalmas,PhysRevB.88.180401,PhysRevB.89.020405,PhysRevLett.98.197203}.  However, there is an abrupt change in shape of the spectra measured at S$\mu$S at $T=105$~K, so it is unlikely that the system enters a magnetic state through a broad ordering transition starting at $T \approx$ 130~K.  Instead we attribute the loss of $A_{\mathrm{rel}}$ on cooling and the rapid relaxation seen in the S$\mu$S data to dynamic fluctuations that have spectral density in the muon time window at temperatures 80$\lesssim T \lesssim$ 130~K.  In the S$\mu$S data, the relaxation rate of the fast-relaxing component at $T=110$~K is $\lambda \approx 300$~MHz, which is sufficiently rapid to dephase this component of the asymmetry at the ISIS time resolution. Approximating the variance of the field distribution by the field measured in the ordered state $\Delta \approx 2\pi\nu(T=0) \approx 2\pi \times100$~MHz, this relaxation rate corresponds to fluctuations with correlation times $\tau \approx 10^{-9}$~s. These fluctuations correspond to energies $\Delta E =\hbar / \tau\approx$~1~$\mu$eV, much lower in energy than the lowest energy magnetic mode ($\approx$ 3 meV) found using inelastic neutron scattering \cite{stock}.   The wide temperature range over which the change in $A_{\mathrm{rel}}$ occurs implies that the muons experience a range of relaxation rates that result in differing fractions of the asymmetry being dephased from the spectra as a function of temperature. Above $T_{\mathrm{N}}$, muons could be subject to magnetic fields with different magnitudes and/or correlation times depending their positions relative to the correlated spins.  Additionally, the morphology of our polycrystalline sample, that likely contributes to the low temperature behaviour discussed below, may also result in muons that occupying sites in differently shaped crystallites experiencing  slightly different relaxation rates.

The drop in $A_{\mathrm{rel}}$ between 130~K and 80~K indicates that 15\% of the asymmetry is due to muons in sites where there are large internal magnetic fields well below $T_\mathrm{N}$.  Below $T_\mathrm{N}$ the only muons that contribute to $A_{\mathrm{rel}}=7$\% are those where there is no large magnetic field field.  These could reasonably be attributed to muons stopping in the sample holder, but might include muons that stop in regions of the sample where there is no strong magnetic field. 
The relaxation rate $\lambda_\mathrm{osc}$ is very small and is consistent with being due to fluctuations in small fields experienced by those muons not subject to the large internal magnetic field. The rate $\lambda_\mathrm{osc}$ shows a maximum slightly above the ordering temperature and a minimum around 80~K.

The baseline asymmetry offset $A_{\mathrm{b}}$ [figure \ref{figure5}(c)] is very small above $T \approx$~140~K. 
It initially increases as temperature decreases down to $T \approx 80$ K, where it reaches a plateau around 4.5\%, accounting, approximately, for those 1/3 of muon spins in magnetic sites aligned along the internal field direction. 
An unusual subsequent drop is seen in $A_\mathrm{b}$ on cooling below 50~K. Given that the precession frequency seen in the S$\mu$S measurements does not show any anomaly in this region, the drop is suggestive of additional dynamics in the field distribution entering the muon time window and relaxing some fraction of the muon spins. 
These fluctuations are distinct from those entering the muon time window above $T_\mathrm{N}$ as they affect only a fraction of the muons and freeze out on further cooling.%This likely reflects the slowing down (or freezing) of fluctuations previously too fast for the muon to respond to and therefore narrowed from the spectra at higher temperatures.

The temperature dependence of the phase $\phi_0$ [figure \ref{figure5}(a), inset] likely reflects the fact that the precessing muons experience small fields whose complicated distribution changes when the system orders.  In addition to the drop in $\phi_0$ close to $T_\mathrm{N}$, we see a large discontinuous change in $\phi_0$ at around 20~K, coincident with the slowing of dynamics and freezing discussed in more detail below.

Further ZF measurements were made at ISIS to investigate the low temperature dynamics.   
Below $T_{\mathrm{N}}$ in these measurements we detect the relaxation of those muons with their spin aligned in the direction of the internal magnetic field (expected to be 1/3 of the total muons for a quasistatically ordered system), that can only be relaxed by dynamic fluctuations. These relaxing spectra change their shape significantly over the temperature regime and so to parametrize them
we fit the asymmetry $A(t)$ to a stretched exponential function, $A(t)=A_\mathrm{rel} e^{-(\lambda_\mathrm{s} t)^{\beta}}+A_\mathrm{bg}$. 

The background asymmetry $A_\mathrm{bg}$ due to muons stopping in nonmagnetic sites is approximately constant and was therefore fixed to the value 3.5\%, estimated from a high-statistics fit. (This value differs from the value of 7\% in the wTF measurements owing to the sample being remounted for the ZF measurements and covering a different fraction of the muon beam profile.) 
The temperature dependence of the relaxing amplitude $A_\mathrm{rel}$, the relaxation rate $\lambda_\mathrm{s}$ and the lineshape parameter $\beta$ is shown in figure~\ref{figure5}.  

We again observe a change in the relaxing asymmetry $A_\mathrm{rel}$ [figure \ref{figure5}(d)] between $T\approx 120~\mathrm{K}$ and $T\approx 80~\mathrm{K}$.  We lose approximately two-thirds of the relaxing asymmetry on crossing $T_\mathrm{N}$ from above, as expected from the arguments above and consistent with the material being magnetically ordered throughout its bulk. 
The lineshape parameter $\beta$ decreases with decreasing temperature across the ordering transition.  Below $T_\mathrm{N}$ the muon experiences approximately exponential relaxation due to electronic moments.  Above $T_\mathrm{N}$ the electronic moments fluctuate very rapidly and are motionally narrowed from the spectra, leaving the muon sensitive to quasistatic nuclear moments which result in approximately Gaussian relaxation.
%and likely reflects the limitations on the ISIS time resolution in capturing the shape of the spectra. 

The relaxation rate $\lambda_\mathrm{s}$ reaches a local maximum just above $T_{\mathrm{N}}$ due to a sudden increase in the magnitude of the local field from the ordering of electronic moments and the critical slowing of fluctuations, both of which contribute to a large relaxation rate.  
It then decreases with decreasing temperature before rising sharply below 40~K, peaking at $\approx$ 20~K and remaining roughly constant at lower temperatures. This is suggestive of fluctuations slowing to be within the muon response time, followed by a subsequent freezing below 20~K.  

\section{Muon stopping site}

\begin{figure*}
	\centering
	\includegraphics[width=\textwidth]{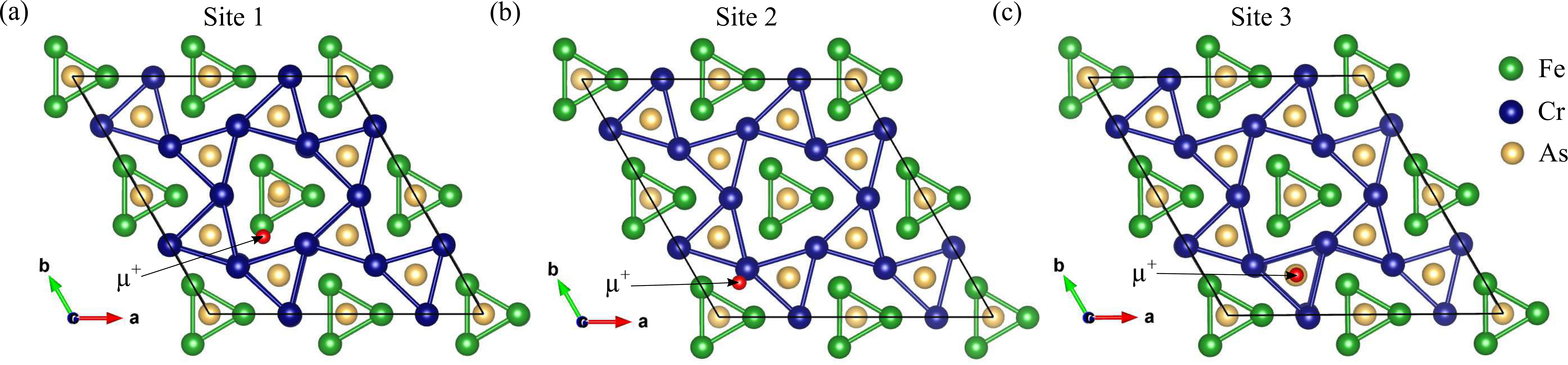}
	\caption{Muon positions within the relaxed structure for each of the stable muon sites within a $2 \times 2 \times 2$ supercell of FeCrAs. Sites shown in (a) and (c) involve a muon sitting in the Cr layer.  For Site 2 shown in (b) the muon sits in the Fe layer.}
	\label{figure6}
\end{figure*}

In order to understand the origin of the contributions to the $\mu^{+}$SR signal,
we carried out density functional theory (DFT) calculations in order to locate the most probable muon stopping sites and assess the degree of perturbation the muon-probe causes in the system. Calculations were carried out using the plane wave based code CASTEP \cite{CASTEP} within the generalized gradient approximation (GGA) \cite{PhysRevLett.77.3865}.  Details of these calculations can be found in the Appendix.  
Structural relaxations suggests the existence of three distinct candidate low-energy stopping sites (see figure~\ref{figure6}) with different energies. These are listed, in order of increasing energy, in table~\ref{table1}.

\begin{table}
  \caption{\label{table1}Positions of the muon sites in FeCrAs and their energies $E$ relative to the most stable site.  Also listed are calculated average dipolar magnetic fields $\nu$ and field distribution widths $\Delta$.}
    % experienced by muons at each site (expressed in terms of the muon precession frequency $\nu=\gamma_{\mu} B / 2\pi$).}
\begin{indented}
\item[]\begin{tabular}{@{}lllll}
\br
 Site & muon position & $E$ (eV) & $\nu$ (MHz) & $\Delta $ (MHz)\\
\mr
1 & Cr layer, hexagon & - & 88 & 44\\
2 & Fe layer & 0.12 & 29 & 9\\
3 & Cr layer, triangle & 0.40 & 50 & 18 \\
\br
\end{tabular}
\end{indented}
\end{table}

Two of the sites involve a muon stopping within the Cr layer: the most stable within a distorted hexagon of the kagome lattice [figure \ref{figure6}(a), lower in energy] and the other inside a triangle in the kagome lattice [figure \ref{figure6}(c), higher in energy].  We also find a stopping site within the Fe layer [figure \ref{figure6}(b)] whose energy lies in between.
These sites are consistent with the minima of electrostatic potential within Cr and Fe layers (see Appendix).  

Our measurements suggest a single site being realized, which we would expect to be the lowest energy one. 
To compare the candidate sites with experiment, we carried out dipolar field calculations (details of which are given in the Appendix).  The non-collinear antiferromagnetic magnetic structure means that each crystallographically distinct site corresponds to a set of magnetically distinct sites, each experiencing a different local magnetic field.  We report the average field and the field distribution width $\Delta ~=~\sqrt{{\gamma}_{\mu}^{2} \langle(B-\langle B \rangle)^{2} \rangle }$
for each set of sites in table~\ref{table1}.
The average dipolar field at the lowest energy site is consistent with the observed precession frequency and so we assign this as the stopping site realized in the material. Displacements of the Cr and Fe atoms due to the presence of a muon at this site are small ($<$0.11~\AA~and $<$0.6~\AA~respectively), with the largest perturbation to the structure being a radial displacement of the nearest As atom of 0.2~\AA.  The lack of any significant structural distortions suggests that the implanted muon should not affect the magnetism measured in this system.

\section{Discussion}
Our $\mu^{+}$SR results indicate that upon cooling below $T \approx 130$~K magnetic fluctations enter the muon time window that persist down to $T \approx 80$~K.  These fluctuations correspond to an energy scale  $E\approx 1~\mu$eV, much lower in energy than the magnetic dynamics measured previously \cite{stock}. Measurements made at S$\mu$S suggest a picture of this material adopting a magnetically ordered state with a magnetic correlation length long enough for coherent muon-spin precession to be resolved below 105~K. The observation of a single oscillatory period in the asymmetry provides an approximate lower bound on this magnetic correlation length of around 10$a$ (where $a$ is the nearest-neighbor spin
separation) \cite{dalmas}. Below $T_\mathrm{N}$ the material is quasi-statically magnetically ordered throughout the whole of its bulk, with no sizeable missing fraction of asymmetry (that could suggest phase separation, for example)
Moreover, our x-ray measurements confirm the existence of ordered antiferromagnetism adopted by Cr moments and no detectable moment on the Fe atoms.  

Evidence for a spin freezing transition at around 20~K is seen in magnetic susceptibility measurements on polycrystalline samples (figure~\ref{figure2}), but in high-quality single crystals a much weaker freezing transition is observed at much lower temperatures (below 10 K).  The behaviour of $A_{\mathrm{b}}$ in the wTF measurements [figure \ref{figure5}(c)] and $\lambda_{\mathrm{s}}$ in the ZF measurements [figure \ref{figure5}(e)] is consistent with this spin freezing transition.  This transition must open additional relaxation channels because it provides a means to relax the 1/3-tail rapidly on the ISIS timescale (thus decreasing $A_{\mathrm{b}}$) with the peak in $\lambda_{\mathrm{s}}$ suggesting a freezing of these dynamics.  Our wTF measurements suggest that the low temperature dynamics do not affect all of the muons.  
The freezing likely results from disorder in the polycrystalline materials that relieves the frustration and which might be expected to result from strain or surface/edge states in the crystallites. We might speculate that the resulting edge states involve either the fluctuating component of the Cr moments, or fluctuating Fe spins whose moments are not reduced to zero by the unusual collective electronic state adopted by the material. Only those muons close to these spins would then experience the glassy freezing of these spin fluctuations. 

In conclusion, an unusual electronic state is adopted by FeCrAs along with a magnetic structure comprising near-zero magnetic moments on the Fe sites. In the magnetically ordered phase, the material is ordered throughout the whole of its bulk.  Low-energy spin fluctuations enter the $\mu^+$SR time window below around 130~K, providing evidence of slow spin dynamics. Polycrystalline samples undergo an additional freezing of dynamics at low temperatures, likely representative of disorder relieving the frustration in the system. 

\ack
Part of this work was carried out at Diamond Light Source, UK, the STFC ISIS facility, UK and the Swiss Muon Source (S$\mu$S), Paul Scherrer Institut, Switzerland.  We acknowledge Diamond Light Source for time on Beamline I16 under Proposal 15119.
We are grateful to Alex Amato for experimental assistance.  We thank EPSRC for financial support under grants EP/N024028/1, EP/N024486/1, EP/N023803/1 and EP/N032128/1. BMH thanks STFC for support via a studentship.  We acknowledge computing resources provided by Durham Hamilton HPC and the UK national high performance computing service ARCHER.  We acknowledge the support of the Natural Sciences and Engineering
Research Council of Canada (NSERC), [funding reference number
RGPIN-2014-04554].  Data presented here will be made available via \url{https://doi.org/10.15128/r2qb98mf48k}.

\appendix
\section{Computational details for DFT calculations}

We used a supercell consisting of $2 \times 2 \times 2$ unit cells in order to minimize the effects of muon self-interaction resulting from the periodic boundary conditions. The system was treated as paramagnetic; spin-polarized calculations are likely to better capture the magnetic state of the system, but computation times were found to be prohibitive.  The structure of the pristine crystal was allowed to relax and the lattice parameters allowed to vary. The experimental and optimized lattice parameters and ionic positions are reported in table \ref{table2}. 

\begin{table}[h]
	\caption{\label{table2}Lattice parameters and ion positions, taken from experiment \cite{hollan,nylund,guerin} and from a DFT calculation (using the PBE functional).}
	\begin{indented}
		\item[]\begin{tabular}{@{}lll}
			\br
			& Experimental & DFT Calculation\\
			\mr
			$a$ (\AA) & 6.096 & 5.905 \\
		$c$ (\AA) & 3.651 & 3.695 \\
		Fe({\it x}\,0$\frac{1}{2}$) & 0.2505 & 0.2502 \\
		Cr({\it x}\,00) & 0.5925 & 0.5838 \\
			\br
		\end{tabular}
	\end{indented}
\end{table}

Muons, modelled by an ultrasoft hydrogen pseudopotential, were placed  in range of low-symmetry positions and the structure was allowed to relax (keeping the unit cell fixed) until the change in energy per ion was less than $2 \times10^{-5}$ eV and the maximum force was below $5 \times 10^{-2}$ eV/\AA.  We used a cutoff energy of 800 eV, resulting in total energies that converge at the order of a few meV per supercell and a $2 \times 2 \times 4$ Monkhurst-Pack grid \cite{MPgrid} for k-point sampling.  

The muon positions within the relaxed structure are shown in figure~\ref{figure6}. In the lowest energy site [figure~\ref{figure6}(a)], a muon sits within one of the distorted hexagons in the Cr layer.
In the second most stable site [figure~\ref{figure6}(b)] the muon sits in the Fe layer.  This site is 0.12 eV higher in energy than the lowest energy site.
The third most stable site [figure~\ref{figure6}(c)] is also in the Cr layer, but here the muon sits in the centre of one of the triangles making up the kagome lattice.
This site is significantly higher in energy than the other two sites, being 0.4~eV higher in energy than the lowest energy site.  For both sites in the Cr layer (sites 1 and 3) the nearest As ion is displaced by $\approx$ 0.2~\AA. Displacements of the Cr and Fe atoms are $<$ 0.11~\AA~for site 1.  The ionic displacements in sites 2 and 3 are generally smaller and are $<$ 0.06~\AA~for Cr and Fe atoms. The muon-induced displacements of the magnetic ions in this system are slightly smaller than those we have previously calculated for the polar magnetic semiconductor GaV$_4$S$_8$ \cite{PhysRevB.98.054428}, with the displacements of the V ions in this system being between $\approx$ 0.1 \AA~and $\approx$ 0.2 \AA~and much smaller than those calculated in the molecular spin ladder compound (Hpip)$_2$CuBr$_4$, where the nearest Cu$^{2+}$ ion is displaced by up to 0.77 \AA~\cite{spinladder}.

\begin{figure}[h]
	\centering
	\includegraphics[width=\columnwidth]{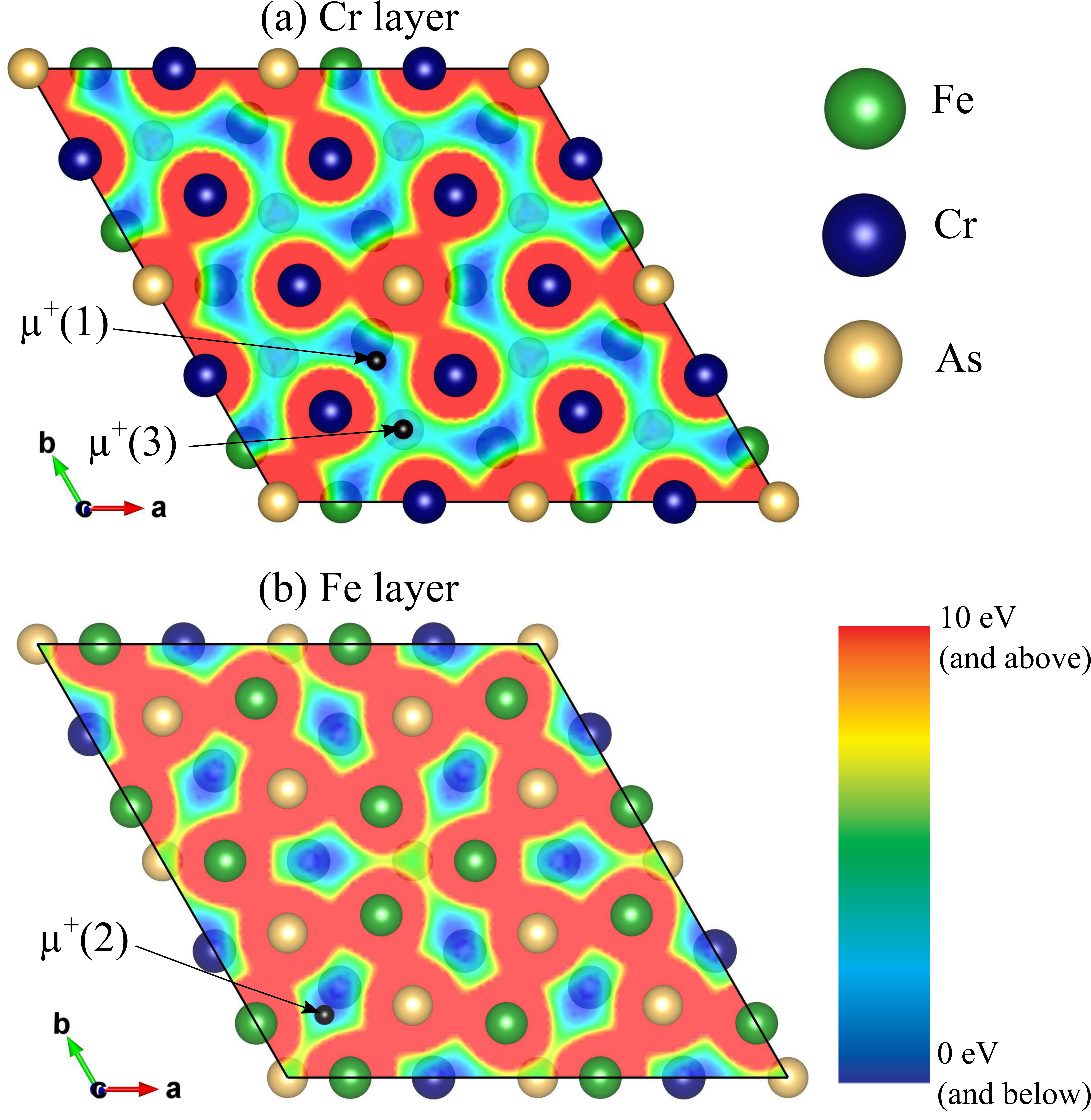}
	\caption{The electrostatic potential within (a) the Cr layer and (b) the Fe layer for FeCrAs.  Blue colouring indicates regions that are attractive to a positive charge, red indicates regions that repel a positive charge. Also indicated are the positions of each of the stable muon sites.  Data are visualized using the VESTA software \cite{vesta}.}
	\label{figurea1}
\end{figure}

In the absence of strong muon-lattice interactions and screening of the $\mu^+$ charge, the minimum of the electrostatic potential provides a good candidate for the muon stopping site \cite{hideseek}.
We have therefore calculated the electrostatic potential for FeCrAs and present this in figure~\ref{figurea1}.  In the Cr layer [figure~\ref{figurea1}(a)], we note that site 1 is slightly displaced from the local minimum, which may be due to the effect of shielding of the muon or from changes to the potential energy landscape due to ionic displacements induced by the muon (though as discussed above, these displacements are very small).  The minimum corresponding to site 1 is seen to be a deeper minimum than the local minimum at site 3, explaining the relative stability of these two sites.  In the Fe layers [figure~\ref{figurea1}(b)] the muon stopping site is also close to the minimum of the electrostatic potential.  From the symmetry of this layer it is clear that all of the minima in the Fe layer are equivalent and we therefore find only one distinct minimum.

\begin{figure}[ht]
	\centering
	\includegraphics[width=\columnwidth]{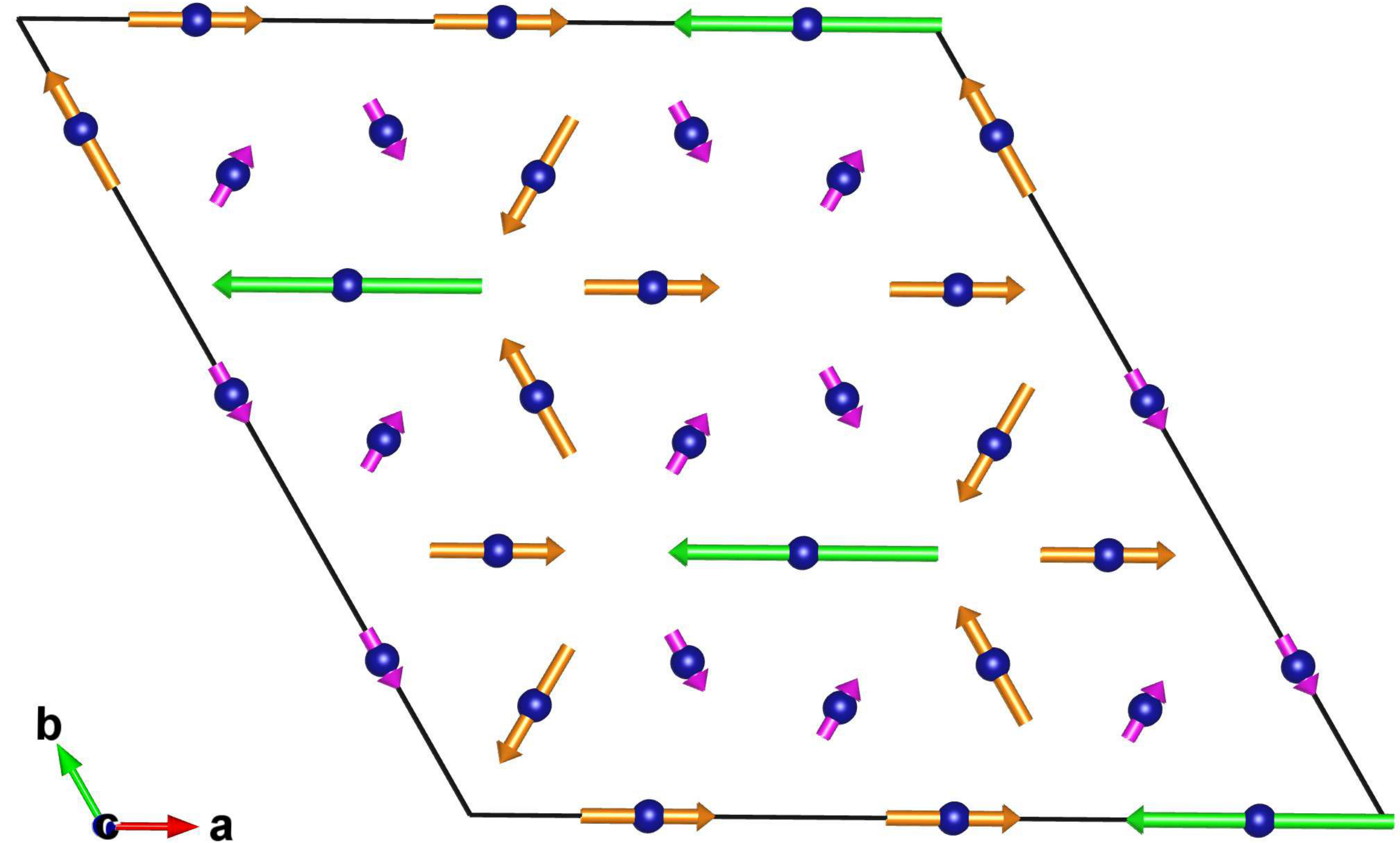}
	\caption{The magnetic structure of the Cr layer in a $3 \times 3 \times 1$ magnetic unit cell. Moments of magnitude 0.635 $\mu_{\mathrm{B}}$, $2 \times$ 0.635 $\mu_{\mathrm{B}}$ and $4 \times$ 0.635 $\mu_{\mathrm{B}}$ are coloured  violet, orange and green respectively. No measurable moments were detected on the Fe sublattice \cite{doi:10.1139/P09-050}.}
	\label{figurea2}
\end{figure}

In order to test the plausibility of the calculated muon sites, we carried out dipolar field calculations.  The magnetic structure of FeCrAs has been determined using neutron scattering \cite{doi:10.1139/P09-050} and is shown in figure~\ref{figurea2}.  A supercell containing $3 \times 3 \times 1$ unit cells is required to describe the non-collinear AFM magnetic structure.

The total magnetic field at the muon site is the sum of dipolar, demagnetizing, Lorentz fields and hyperfine interactions.  However, as FeCrAs orders antiferromagnetically, the demagnetizing and Lorentz field are zero.  We therefore take the dipolar coupling to be the dominant contribution to the field at the muon site.  The dipolar field experienced by a muon at position $\textbf{r}_{\mu}$ is given by

\begin{equation}
\textbf{B}_{\mathrm{dipole}}(\textbf{r}_{\mu})=\sum_{i} \frac{\mu_0}{4 \pi r^3} \left[ 3(\boldsymbol{\mu}_i \cdot \hat{\textbf{r}}) \hat{\textbf{r}} - \boldsymbol{\mu}_i  \right],
\label{equationa1}
\end{equation}
where $\mu_0$ is the permeability of free space and $\textbf{r}=\textbf{r}_{\mu}-\textbf{r}_i$ is the position of the muon relative to ion $i$ with magnetic moment $\boldsymbol{\mu}_i$.  By considering the moments on the Cr sublattice only, we calculate the magnitude of the dipolar field (and therefore the expected muon precession frequency) at each of the muon stopping sites.  The magnetic unit cell (MUC) differs from the crystallographic unit cell and therefore sites that are crystallographically equivalent are not necessarily magnetically equivalent.  For each distinct muon stopping site we generate all of the crystallographically equivalent positions within the MUC.  Computing the dipolar field associated with each of these positions allows us to determine the average field and field distribution width, $\Delta =\gamma_{\mu} \sqrt{{\langle B-\langle B \rangle \rangle}^2}$, associated with each of the classes of muon site, which we report in table~\ref{table1} in the main text.  Note that these fields have been calculated for the undistorted structure, though as we have noted earlier, the displacements of the Cr ions due to implanted muon are very small.

\end{document}